\documentclass[article,onecolumn]{aa}

\usepackage[T2A]{fontenc}
\usepackage{graphicx}

\newcommand{\arep}{Astronomy Reports }
\newcommand{\alet}{Astronomy Letters }

\def\bsao{{Bull. Spec. Astrophys. Obs.}}
\def\ab{{Astrophys. Bull., }}

\begin{document}

\title{Detailed spectroscopy of post-AGB supergiant GSC\,04050$-$02366 in IRAS\,Z02229+6208 IR source  system}
\author{V.G.~Klochkova,  V.E.~Panchuk,}
\institute{Special  Astrophysical Observatory, RAS, Nizhnij Arkhyz, 369167 Russia,   \\
   \email{Valentina.R11@yandex.ru}   }

\titlerunning{Detailed spectroscopy of post-AGB supergiant GSC\,04050$-$02366}
\authorrunning{Klochkova \& Panchuk}

\abstract{In the optical spectra of the cold post-AGB supergiant GSC\,04050$-$02366,
obtained with the 6-meter BTA telescope with a spectral resolution of R$\ge$60\,000
on arbitrary dates over 2019$\div$2021, a radial velocity variability is found.
Heliocentric Vr based on the positional measurements  of numerous absorptions
varies from date to date with a standard deviation of $\Delta$Vr$\approx$1.4\,km/s
about the average value of Vr=24.75\,km/s, which may stem out of the low-amplitude
pulsations in the atmosphere.
The spectra of the star are purely absorption type, there are no obvious emissions.
Intensity variability of most of absorptions and  Swan bands of the C$_2$ molecule 
was discovered. A slight asymmetry of the H$\alpha$ profile is observed at some 
observation dates.
The position of H$\alpha$ absorption core varies within 27.3$\div$30.6\,km/s.
Splitting into two components (or asymmetry) of strong low-excitation absorptions
(YII, ZrII, BaII, LaII, CeII, NdII) was found. The position of
the long-wavelength component coincides with the position of other photospheric 
absorptions,  which confirms its formation in the atmosphere of the star. 
The position of the shortwave component is close to the position of the 
rotational features of  Swan bands, which indicates its formation in the 
circumstellar envelope expanding at a velocity of about Vexp=16\,km/s.}

\maketitle
\section{Introduction}\label{intro}

In recent decades, a detailed spectroscopy of by far evolved stars that have 
passed evolutionary stages with nucleosynthesis in the stellar cores and 
episodes with mass loss due to the stellar wind and ejection of the envelope
is carried out at the 6-meter BTA telescope of the Special Astrophysical Observatory RAS.
Numerous results obtained within the framework of the program are briefly presented 
in the survey \citet{SciNES}.
One of the main areas of our research is spectroscopy of massive
hypergiants with initial masses of 20$-$40\,$M_{\odot}$ near the
luminosity limit on the Hertzsprung--Russell diagram. Results of spectral 
monitoring at the BTA of these extremely rare objects are summarized in a recent 
survey \citet{YHG}.
Less massive evolved far stars with initial masses of around 2$-$8~$M_{\odot}$
are observed on the asymptotic giant branch (hereinafter, AGB) in the form of red
supergiants with an effective temperature of T$_{\rm eff}\approx$3000--4500~K.
By far evolved AGB stars at the stages of hydrogen and helium layer  burning
experience a significant mass loss due to the forceful (the rate in
the interval of \mbox{$10^{-8}\div10^{-4}M_\odot$/year})
stellar wind \citep{Hofner}.
Owing to the wind, an expanding envelope of gas and dust is formed around the star,
the presence of which manifests itself in the excess IR flux, specific features in the 
radio, IR and optical spectra. Circumstellar gas and dust envelopes formed at the RGB, 
AGB, and post-AGB stages of evolution appear in large excesses of IR radiation, 
which serves as the main selection criterion for these objects \citep{Kwok93}.

Interest in AGB stars and their nearest descendants---post-AGB stars
in particular stems from the fact that it is in these stars, being in
the short-term evolutionary stage, that physical conditions arise for the synthesis
of nuclei of heavy metals and their subsequent dredge-up into the stellar atmosphere. 
Owing to these processes AGB stars are the main suppliers (over $50$\%) of all
elements heavier than iron, synthesized as a result of the $s$-process, the essence 
of which consists in a slow (compared to $\beta$-decay) neutronization of nuclei.
The main features of stellar evolution near the AGB and the results of up-to-date 
calculations of synthesis and dredge-up of elements are given in the papers by
 \citet{Block1995, Herwig2005, Criscienzo, Liu}.

A part of AGB and post-AGB supergiants is accessible to optical spectroscopy with   
high spectral resolution. We have published the surveys of these studies performed 
on the 6-meter telescope in the papers \citet{rev1, Envelop, rev2}. This program of
observations also includes a cool supergiant GSC\,04050$-$02366, associated
with an IR source IRAS\,Z02229+6208 (hereinafter~ IRAS\,Z02229).  This star of
spectral type G8--K0\,Ia \citep{Hrivnak99} is faint in the visible range
due to a significant interstellar and circumstellar extinction, hence it has 
a short and uncomprehensive history of study.

The data of the InfraRed Astronomical Satellite (IRAS) indicated the presence of 
a powerful flux at the  25\,$\mu$m band from the  IRAS\,Z02229 source. This has
stimulated the beginning of an active study of the object.
1999 turned out to be especially fruitful in the study of the IRAS\,Z02229 system. 
\citet{Hrivnak99}, using low-resolution optical spectra revealed
the presence of absorption bands of the C$_2$ and C$_3$ molecules and assigned the 
object to a scanty group of protoplanetary nebulae (PPN) with an emission at the 
wavelength of $\lambda$=21~$\mu$m.
The same \citep{Hrivnak99} authors summarized the  information about the
IRAS\,Z02229  system, published by the time of their study: a two-humped shape 
of SED,  a color excess $E(B-V)=1.0$, a considerable expansion of the circumstellar
envelope, including its expanse in H$\alpha$; an intense envelope emission in the
$(3;2)$ band of the CO molecule, which made it possible to determine the systemic 
velocity of this object Vsys=+24.3\,km/s and the envelope expansion rate
Vexp=+15.2\,km/s. \linebreak
The main data required for our work, published on the star GSC\,04050$-$02366,
associated with the IR source IRAS\,Z02229, are listed in Table~\ref{star}.
With such a high color excess, the object, the predecessor of GSC\,04050$-$02366
at the previous AGB stage, refers to the so-called ERO C-rich stars,
according to the classification of stars with mighty envelopes (Extremely Red Objects) 
introduced by \citet{Groen}. It is important to note here a significant difference 
between the effective temperature and color excess values for this star, determined 
by various methods: the method of atmospheric models in \citet{Reddy99} and
Spectral Energy Distribution (SED) modeling using the Gaia EDR3 parallaxes \citep{Kamath}.
\begin{table*}[h!]
\smallskip
\caption{Main data about the post-AGB star GSC\,04050$-$02366 associated with the IR source IRAS\,Z02229}
\begin{tabular}{l r  }
\hline
Parameter & \multicolumn{1}{c}{Value} \\
\hline
$\alpha$ & 02$^h$ 26$^m$ 41.8$^s$  \\
$\delta$ & +62\degr 21$^`$ 22\arcsec \\
l/b      & 133.7/1.5\degr    \\
$\pi$    & 0.3806\,mas~\cite{Gaia}    \\
Sp       & G8-K0\,Ia:~\cite{Hrivnak99}\\ 
L/L$_\odot$ & 12959~\cite{Kamath} \\  
T$_{\rm eff}$ & $\approx$5400\,K~\cite{Reddy99} \\   
         & 5952\,K~\cite{Kamath} \\  
E(B-V)   & 1.0~\cite{Hrivnak99} \\ 
         & 1.9~\cite{Kamath} \\  
$\rm [Fe/H]$& $-0.4$~\cite{Reddy99} \\
\hline
\end{tabular}
\label{star}
\end{table*}
The next important step of  GSC\,04050$-$02366 research  was the widely cited
work of Reddy et al. \citet{Reddy99}. These authors, based on a high-resolution 
optical spectrum,  obtained with the 2.7-m telescope echelle spectrograph of 
the McDonald observatory, determined the parameters of the stellar atmosphere 
model and its detailed chemical composition.
As a result,  \citet{Reddy99} came to a conclusion that metallicity of the stellar 
atmosphere is reduced:   ${\rm [Fe/H]_{\odot}}=-0.5$ at large excesses of carbon
${\rm [C/Fe]_{\odot}}=+0.8$ and heavy metals of the $s$-process ${\rm [s/Fe]_{\odot}}=+1.4$.
Such peculiarities  of the chemical composition indicate that the star has passed 
the evolutionary AGB stage and the 3-d dredge-up episode.

So far, there has been no further of the optical spectroscopy  of
GSC\,04050$-$02366, which prompted us to start spectral monitoring of the star 
with high spectral resolution. In this paper
we present the results of the analysis of several optical spectra of GSC\,04050$-$02366,
obtained with the 6-m BTA telescope on arbitrary dates over 2019$\div$2021.
The primary purpose of our work is to search for the possible variability
of the spectral feature profiles  and studying the temporal behavior of the Vr pattern.
Section~\ref{obs} of the paper briefly describes techniques of observation and data analysis.
Section~\ref{results} shows our results in comparison with those published
earlier, in Section~\ref{discuss} we present a discussion of the results obtained, 
while the main conclusions are listed in Section~\ref{conclus}.
\begin{table*}[ht!]
\smallskip
\caption{The results of measurements of the heliocentric radial velocity Vr in the spectra
      of  GSC\,04050$-$02366}
\begin{tabular}{ c| l|  c|  c| c  }
\hline
Date/JD & \multicolumn{4}{c}{\small Vr, km/s} \\  
\cline{2-5}
    &Absorptions & H$\alpha$(core) & Swan & DIBs   \\  
\hline
   1    & \hspace{7mm}  2       &  3                &   4  &  5    \\
\hline
19-23.12.1996$^1$ & 18$\pm 1$ &&6 & \\
\hline
06.12.2019 &20.37$\pm 0.01$\,(283)&32.2&8.8$\pm 0.20$\,(24)  &$-7.4\pm0.5$\,(7) \\
2458824.3  &&&& \\
\hline
26.10.2020 &25.62$\pm 0.01$\,(327)&27.9 &8.6$\pm 0.01$\,(87) & $-6.2\pm0.4$\,(9)\\
2459149.3  &&&& \\
\hline
29.11.2020 &25.81$\pm 0.01$\,(330)&29.9 &9.0$\pm0.10$\,(11) & $-6.7\pm0.7$\,(3)\\
2459182.5  &&&& \\
\hline
23.10.2021 &26.10$\pm 0.01$\,(466)&26.1 & 9.7$\pm0.16$\,(29)& \\
2459511.2  &&&& \\
\hline
\multicolumn{5}{l}{\footnotesize 1 -- mean  Vr for 1996 based on the data from \citet{Reddy99}} \\[-10pt]
\end{tabular}
\label{velocity}
\end{table*}

\section{Observational material and its reduction}\label{obs}

The spectra of GSC\,04050$-$02366 were obtained with the NES echelle spectrograph
 \citep{NES}, stationary located at the Nasmyth focus of the 6-meter BTA telescope. 
The moments of observation of the star are listed in Table~\ref{velocity}.
At present, the NES spectrograph is fitted up with a large-format CCD
with 4608$\times$2048 elements sized 0.0135$\times$0.0135~mm,
readout noise 1.8\,e$^-$.
The registered spectral range  in our  spectra is $\Delta\lambda=470-778$\,nm.
To reduce flux losses not falling  of spectral resolution, the NES spectrograph 
is equipped with an image slicer with a configuration for 3~slices.
Spectral resolution is $\lambda/\Delta\lambda\ge$60\,000,
the ratio of signal to the level of  noise along the echelle order in the 
spectra varies from~40 to~70.
Moreover, the signal is significantly reduced in the short-wavelength part of the 
echelle frame due to a significant absorption of stellar radiation in the 
circumstellar medium.

\section{Spectral features and radial velocity behaviour}\label{results}
To identify the features in the spectra of  GSC\,04050$-$02366 we used the spectral
atlas \citep{atlas2, atlas1} based on detailed spectroscopy of a related
post-AGB star HD\,56126  (IRAS\,07134+1005), as well as the spectral atlas \citep{atlas_Barn}
for carbon stars.

\begin{figure*} \vspace{2mm}
\includegraphics[angle=0,width=0.75\textwidth,bb=40 45 720 520,clip]{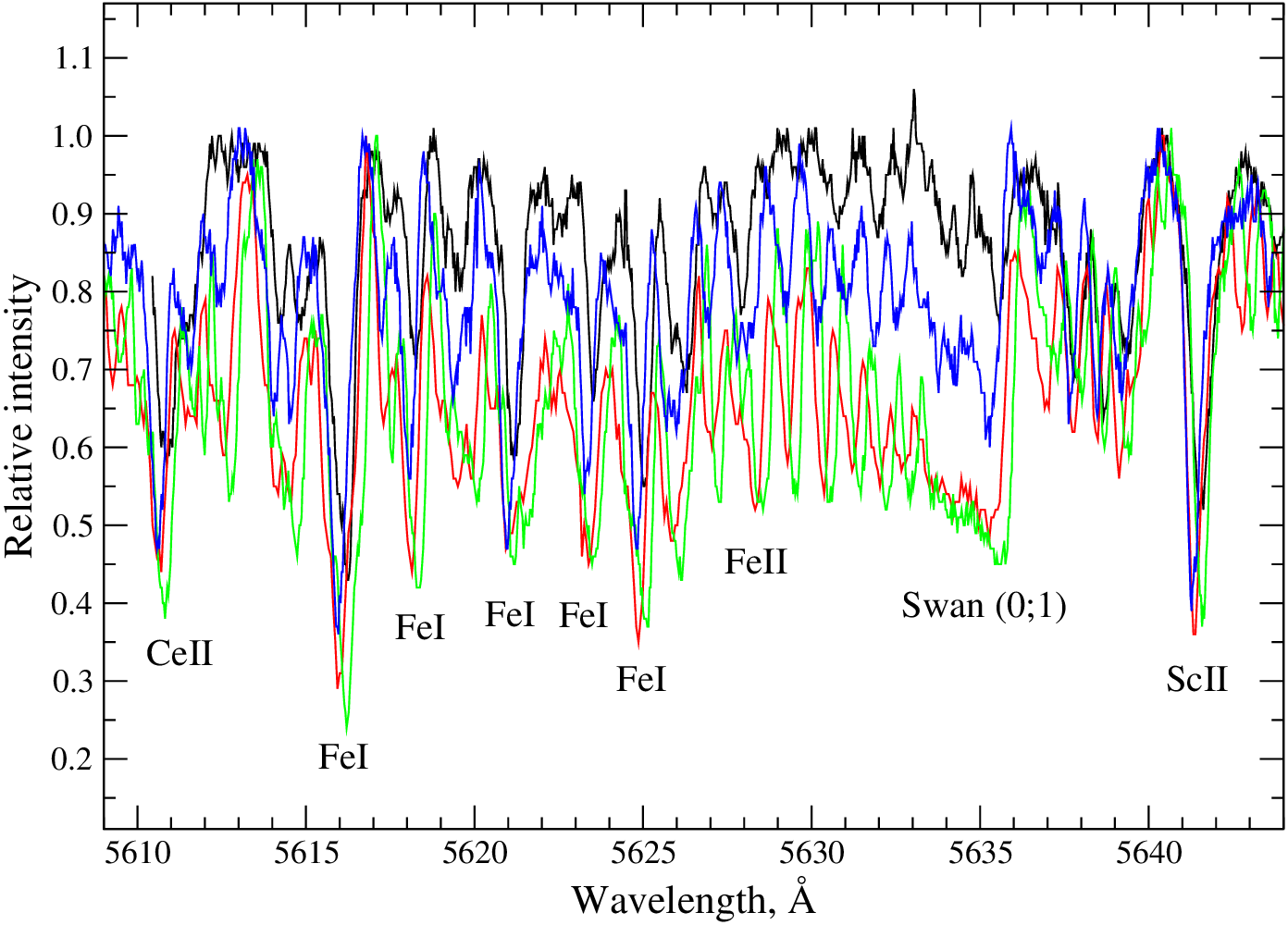}  
    \caption{Fragments of the   spectrum of GSC\,04050$-$02366
    with the C$_2$\,($0;1$) Swan band  registered during different
    observational sets: black line---2019-12-06, red line---2020-10-26,
    green---2020-11-29, blue---2021-10-23. Identification of the main
    absorptions in this fragment, including the edge of the C$_2$ molecule 
   band head at the wavelength   near $\lambda\,5635$\,\AA.}
    \label{Swan}
\end{figure*}

\subsection{Radial velocity}

All results of  heliocentric velocity Vr measurements are given in columns 
$(2)$--$(5)$ of Table~\ref{velocity}.
The average  Vr values  for symmetrical absorptions, rotational components 
of Swan bands and interstellar features of the DIBs  are given
for the samples of details and the errors of these means.

From the comparison of high-accuracy data in the second column of the Table,
we obtain the radial velocity variability: the average  Vr value over
the measurements of positions of many metal absorptions varies from date to date
around the mean value Vr=24.75\,km/s  with the standard deviation
from the mean $\Delta$Vr=1.4\,km/s, which may be a consequence of low-amplitude
pulsations in the atmosphere.

\subsection{Swan bands}

As shown in \citet{Hrivnak99, Reddy99}, one of the most noticeable features 
of the optical spectrum of this star are the bands of carbon-containing  CN, C$_2$ 
and C$_3$ molecules. All our spectra also contain the Swan bands of the C$_2$ molecule.
Moreover, in the spectra we recorded over 2019$-$2021, all Swan bands are present 
as absorptions.
For illustration, Fig.~\ref{Swan} shows a fragment of the spectrum, including 
the Swan band molecules C$_2$\,($0;1$) with the head on $\lambda$=5635\,\AA\ for 
four moments of observations. The variation in the intensity of the band cannot 
be explained by the difference in the observation conditions,
since the spectra of 2019--2020 are obtained on the observation nights with
excellent transparency and seeing $\beta=1\arcsec \div 1\farcs5$. Only at one 
night of 23.10.2021 the seeing was slightly  worse:~$\beta\approx 3\arcsec$.

The variability of the band's intensity in this figure can be explained by
non-sphericity of the envelope, which causes different extents of its contribution 
to the observed spectrum.
This assumption is consistent with the ``elongated'' morphology type for the image of
IRAS\,Z02229 obtained with the HST \citep{Sahai}. Nonsphericity of the dust envelope
in the IRAS\,Z02229 system is also confirmed by a high level of polarization 
\citep{Ueta1, Ueta2, Akras}. The influence of a powerful envelope on the observed 
spectrum of the star as a whole is manifested in such a way.

\subsection{H$\alpha$ profile}

\begin{figure*} \vspace{2mm}
\includegraphics[angle=0,width=0.505\textwidth,bb=10 80 550 680,clip]{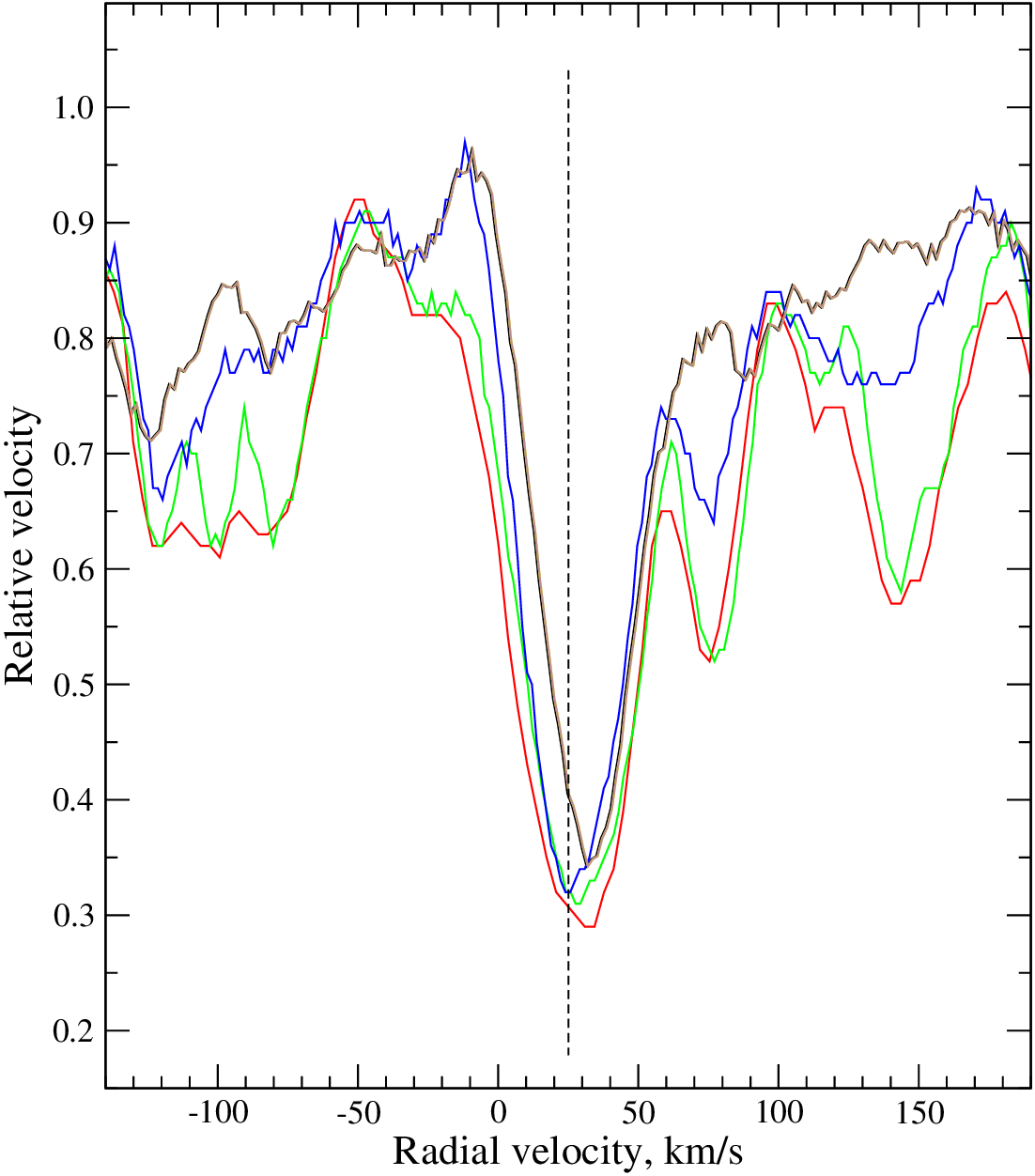}
\caption{H$\alpha$ profile in the ``radial velocity--relative intensity'' coordinates
in the spectra of GSC\,04050$--$02366 obtained on different dates
(the colors of the lines as in Fig.~\ref{Swan}). Here and further
 the position of the dashed vertical line coincides with the value of the systemic velocity
Vsys=+24.3~km/s according to \citet{Hrivnak99}. }
\label{Halpha}
\end{figure*}

The spectra of GSC\,04050$-$02366 do not contain explicit emission features or wind
components typical of early post-AGB stars.
As seen in Fig.~\ref{Halpha}, the H$\alpha$ profile in the GSC\,04050$-$02366 spectra is
almost symmetrical, only at some dates there is a slight asymmetry of the profile
in the form of an elevation of its short-wave wing. Moreover, this is observed only for
those two observation time instants for which the Swan band is significantly weakened
in Fig.~\ref{Swan}. As follows from the data in the third column of the Table~\ref{velocity},
the position of the H$\alpha$  absorption core varies within 27.3$-$30.6\,km/s.
For most of the dates, the core is redshifted by about 2\,km/s.
This difference, according to the data of \citet{Reddy99}, is many times higher
in the $1996$ spectrum.

\begin{figure*}[]
\includegraphics[angle=0,width=0.45\textwidth,bb=20 80 550 680,clip]{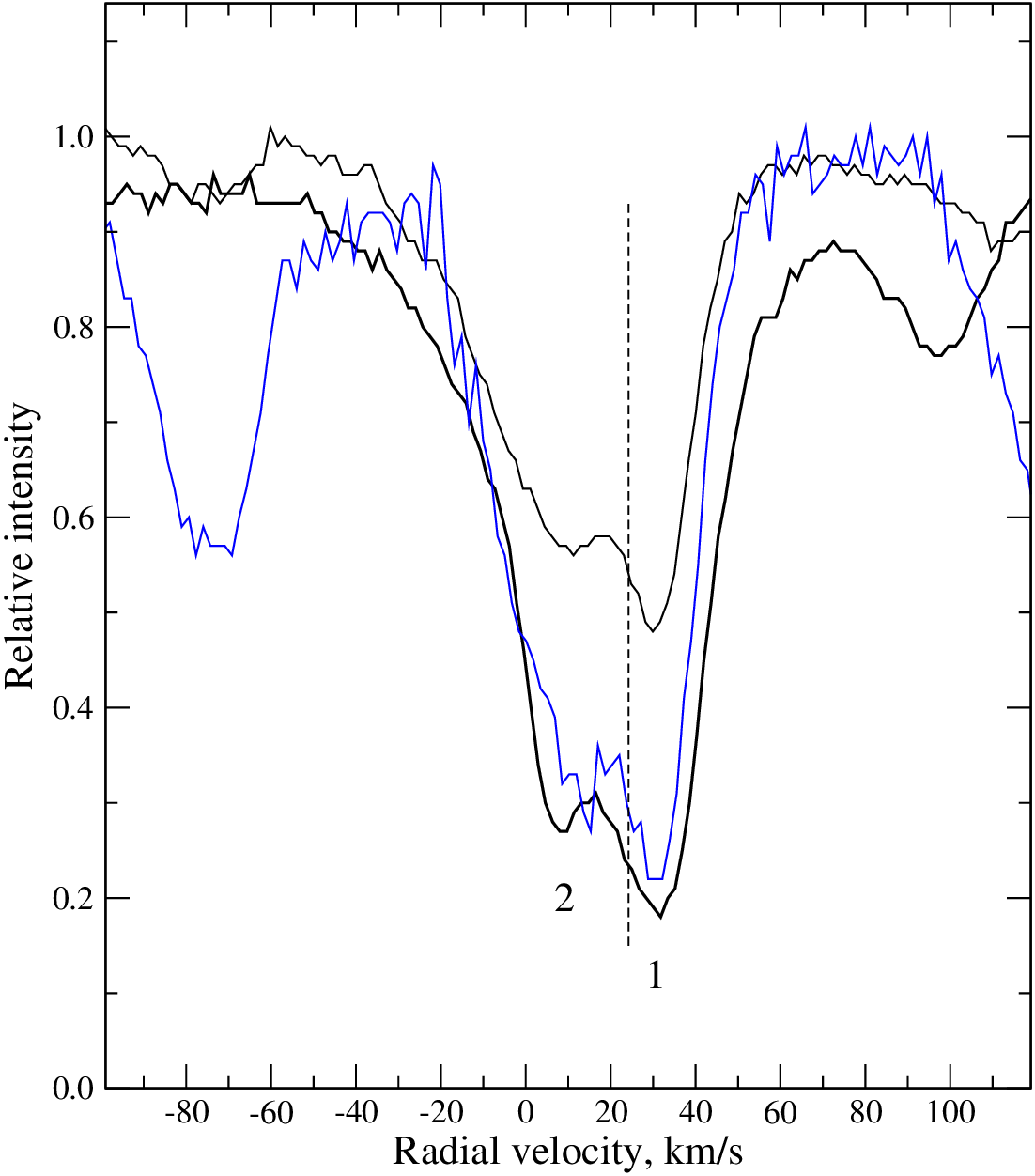}
    \caption{The profiles of BaII\,$\lambda\,6141$\,\AA\ (thick black line), YII\,$\lambda\,5200$\,\AA\
    (blue line) and LaII\,$\lambda\,6390$\,\AA\ (thin black line) absorptions in the spectrum for 2019-12-06.
    The ``1'' component, the position of which is consistent with the position of the photospheric absorptions,
    and the short-wavelength component ``2'', which is a circumstellar component are marked (for more 
    details, see the text).}
    \label{BaLaY_s693}
\end{figure*}

In the spectra of the cool supergiant with a slight metallicity deficit ([Fe/H]=$-0.5$ \citep{Reddy99}) 
the absorptions of rotational transitions in the Swan system bands are mostly blended. 
However, the high spectral resolution of our data allows to single out selected features 
of the rotational transitions of the Swan bands,  focusing on their profiles, which are more 
narrow  compared to the photospheric absorptions.
The average values  of the measured velocities Vr(Swan), their errors and the number of used
features are given in the fourth column of the Table~\ref{velocity}.

\begin{figure*}[] \vspace{2mm}
\includegraphics[angle=0,width=0.4\columnwidth,bb=20 70 550 680,clip]{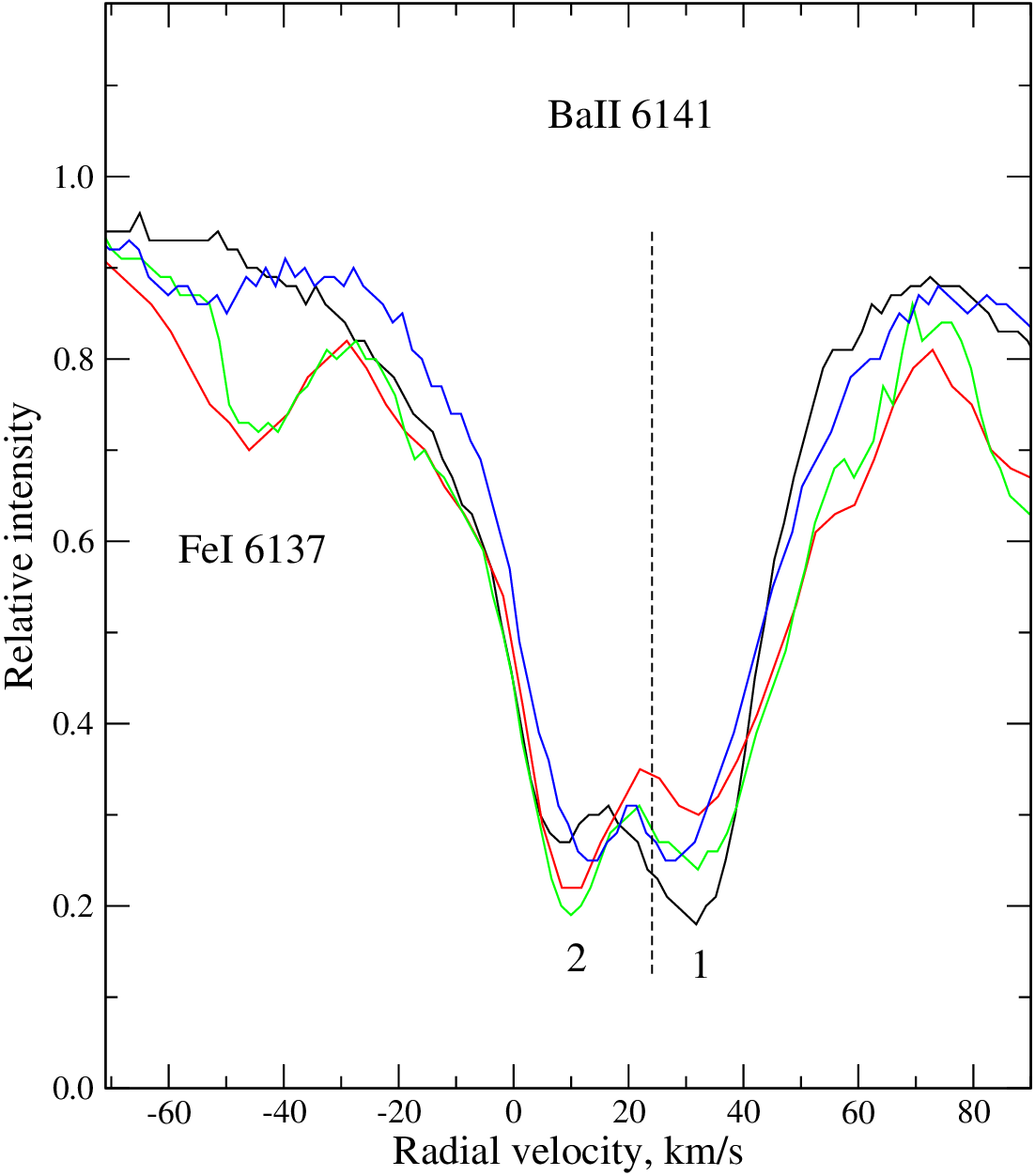}
\includegraphics[angle=0,width=0.4\columnwidth,bb=20 70 550 680,clip]{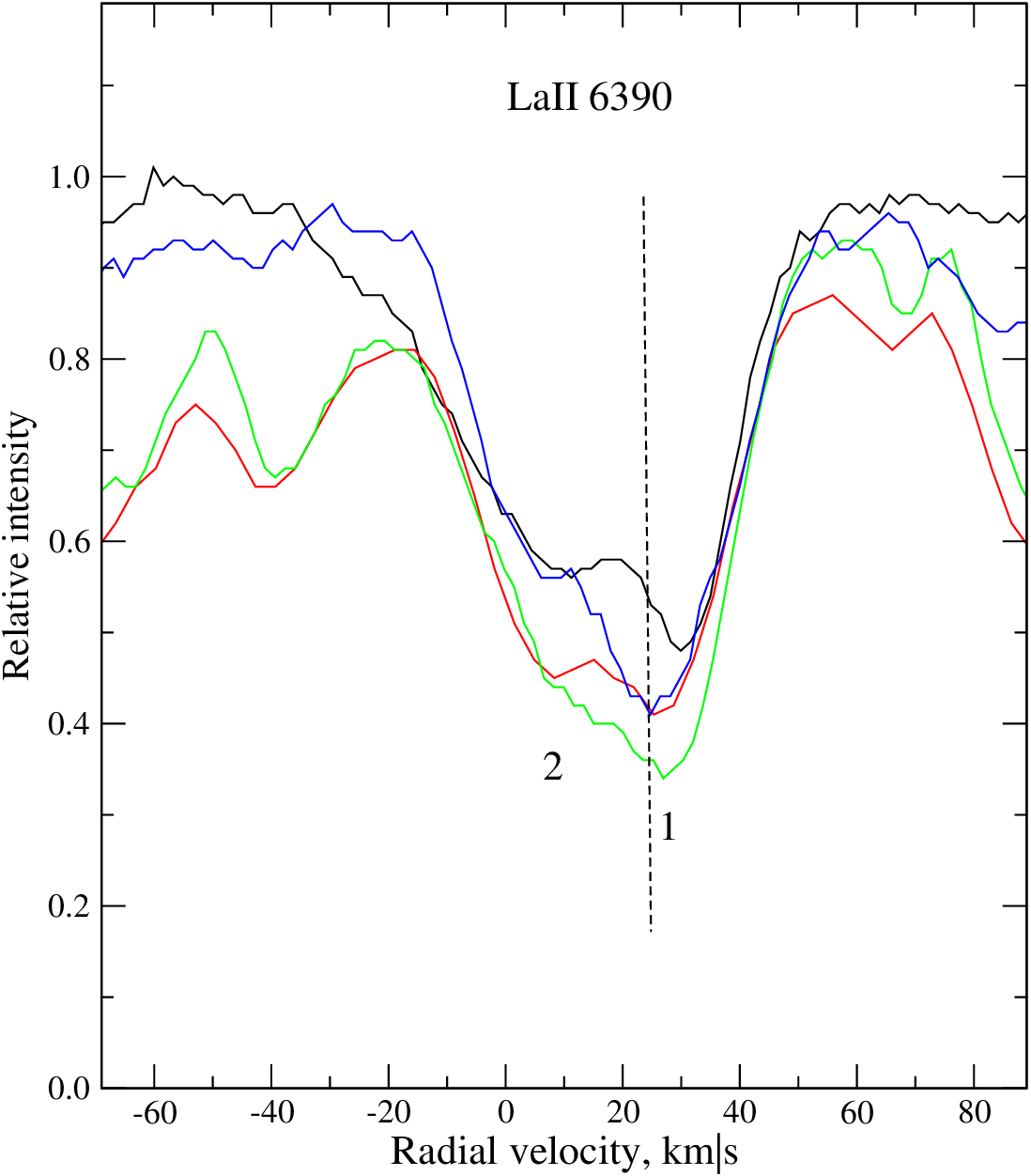}
\caption{Same as in Fig.~\ref{Halpha}, but for the  BaII\,6141 and
      LaII\,6390 absorptions in the spectra for four dates.}
\label{BaLa_4}
\end{figure*}

\subsection{Split absorptions}

\citet{Reddy99} noted that intensity of the BaII ion lines in the GSC\,04050$-$02366 spectrum
is too high, which makes it difficult to model them and calculate the barium abundance 
in the atmosphere of the star. Figure~\ref{BaLaY_s693} demonstrates the BaII\,6141 line profile
in our spectrum for 06.12.2019.  Here you can clearly see the complex structure of the 
profile of this line, split into two components.
The ``1'' component is marked in this figure, the position of which is consistent with the 
position of other photospheric absorptions, which indicates its formation in the atmosphere 
of the star. The shortwave component ``2'' is an envelope component, since its position 
corresponds to the velocity of Vr$\approx$9.0\,km/s close to the Vr calculated from the
positions of rotational features of  Swan bands.
The same figure additionally shows the profiles of strong absorptions
LaII\,6390  and YII\,5200  split into similar components.  As you can see in Fig.~\ref{BaLaY_s693}, 
the width of the shortwave component of each line is significantly lower than the one 
of the long-wavelength one, which confirms formation of the short-wavelength component 
in the envelope.


In the spectra of GSC\,04050$-$02366, the strongest absorptions of metals or ions
(KI, NaI, YII, ZrII, BaII, LaII, CeII, NdII) with a lower energy level
low excitation potential (\mbox{$\chi_{\rm low}\le 1$~eV}) are split into 
two components or have asymmetrical profiles.
In the presence of an explicit splitting, the position of their long-wavelength component
coincides with the position of unsplit absorptions of other metals, which confirms
its formation in the atmosphere of the star. Position of the shortwave component
is close to the position of Swan bands features, which reflects its formation in
the circumstellar environment.  This splitting effect obviously increases the equivalent 
width of the corresponding absorptions and given a reduced spectral resolution 
it has to be taken into account in determining the abundance of heavy metals.

Figure~\ref{BaLa_4} compares the profiles of the split  BaII\,6141 and
LaII\,6390 profiles in the spectra over four dates of our observations.
In general, the character of  splitting is the same  for all observation 
dates.

\begin{figure*}[] \vspace{2mm}
\includegraphics[angle=0,width=0.4\textwidth,bb=20 70 545 675,clip]{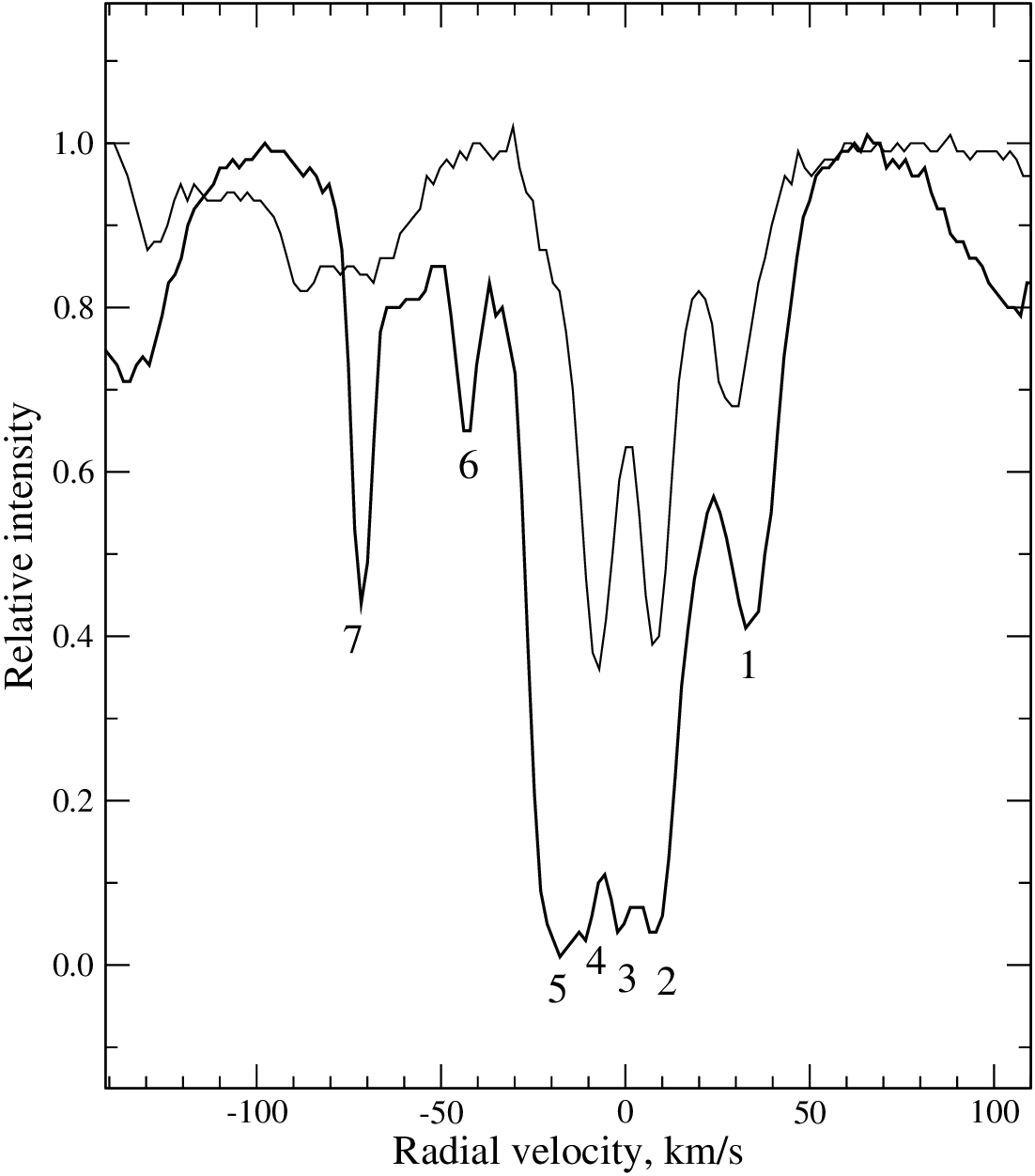}
\caption{Multicomponent  NaI\,5895 (thick line) and KI\,7699  line profiles in the spectrum of
  GSC\,04050$-$02366 obtained on 2019-12-06.  The NaI line profile components marked: ``1''
  and ``2'' are formed in the stellar  atmosphere and in its circumstellar envelope, respectively,
  while the ``3''--``7''  components are interstellar.}
\label{Na_K}
\end{figure*}

\subsection{Interstellar features}

In the spectrum of a distant star near the Galactic plane we naturally expect 
to find the presence of a set of interstellar features, which is rather well 
illustrated by  the profiles of the  NaI\,5895  (thick line) and KI\,7699 lines 
in Fig.~\ref{Na_K}.
Here, the component ``1'' (Vr$\approx30$~km/s) is formed in the stellar atmosphere, 
the component ``2''  (Vr$\approx7.4$~km/s)---in the circumstellar envelope, where 
the Swan bands are also formed.  Only the  ``6'' and ``7'' components are reliably 
identified in the NaI\,5895 line profile.
The component Vr=$-7.5$~km/s is present in the KI\,7699 line profile, its interstellar 
origin is confirmed by the agreement of this velocity with the velocities in the 
last column of Table~\ref{velocity} by the identified DIBs.
The velocity range from Vr=$-2.0$\,km/s to Vr=$-22.0$\,km/s,
which contains the ``3--5'' components of the  NaI profile, is a set of  interstellar 
absorptions, not resolvable at the spectral resolution of the NES spectrograph.
It should also be taken into account that during long exposures of a faint star,
a  NaI emission of telluric origin may be registered, the position of which
in the heliocentric system of velocities varies from date to date of observations.

Let us mention that the close interstellar components  (Vr=$-8.5,-44.4,-65.3$\,km/s)
of the NaI\,5889 and 5895 lines  profile  have been identified earlier in the spectra of
the central star of the IRAS\,02441+6922 \citep{RAFGL5081}, whose galactic coordinates
${l/b}=132\fdg9/9\fdg2$ differ little from the coordinates of IRAS\,Z02229.
In the spectrum of this object, the DIB sample positions were reliably measured:
their position corresponds to the velocity of  Vr=$-9.3\pm0.3$\,km/s.
Interstellar components of the NaI doublet with close velocities were registered by
\citet{VES723} and in the spectrum of a distant hot star VES\,723, whose galactic 
coordinates are also close to the coordinates of   GSC\,04050$-$02366:
${l/b}=132\fdg4/1\fdg1$.

It is difficult to identify the diffuse bands of DIBs in the spectra of the 
cool star with a rich absorption spectrum combined with molecular bands.
In addition to the strongest DIBs at $\lambda\,5780$\,\AA\ and $\lambda\,6284$\,\AA,
we managed to identify only a few DIBs whose average position, Vr$\approx-6.7$\,km/s,
agrees with the position of the interstellar component KI.

To conclude this Section, let us note that  \citet{Reyniers} showed
that the absorption at $\lambda\approx6707.8$\,\AA\ should be identified with
CeII\,$\lambda\,6708.099$\,\AA{}.  For example, the authors of \citet{Reddy99, V2324Cyg} 
identified this absorption in the spectra of post-AGB stars with the  LiI\,$\lambda\,6707.76$\,\AA\ 
line and concluded that there is an excess of lithium in the atmospheres of these stars.
\citet{Reddy99} noted   unreliability of finding the lithium abundance  in the atmosphere of 
GSC\,04050$-$02366 based on absorption intensity near $\lambda\,6707$\,\AA\ due to the
possible blending by the CN and Ce lines.  We have reliably
identified this absorption with the  CeII\,$\lambda\,6708.099$\,\AA{} line, 
since the radial velocity measurements based on the position of this feature
are in excellent agreement with the mean Vr from the metal lines.

\section{Discussion of the results}\label{discuss}
All the GSC\,04050$-$02366 spectra we have are purely absorption spectra.
At the same time, we have revealed the variability of the spectrum and the
peculiarities of the behavior of a number of absorptions.  Photometric observations 
of this star also indicate a significant variability.  According to the  observations 
of GSC\,04050$-$02366 in  2014$-$2018 from the ASAS-SN survey of \citep{Kochanek},
the apparent brightness of the star varies in the range of \mbox{$11\fm6$--$12\fm2$}.
Based on the analysis of these long observations,  the period  P=443$\fd04$
and variability amplitude $\Delta V=0\fm47$  estimates are made in the ASAS-SN database.
Somewhat earlier, the photometric features of GSC\,04050$-$02366 among the sample of
related post-AGB stars were studied by \citet{Hrivnak2010}.
These authors revealed a long-term brightness variability, the presence of several 
periods with a dominant one of around $153$~days, as well as a combination of deep minima
(up to $\Delta V \approx 0\fm54$) with flatter ones.  The set of photometric
data for IRAS\,Z02229 and related objects allowed these authors to
make a choice in favor of pulsations, and not binarity, as the cause of variability.

There are no wind components in the  GSC\,04050$-$02366 spectra and
emissions  typical for the spectra of post-AGB stars, occurring in 
the atmosphere of a star or in its circumstellar environment.
For example, in the spectrum of HD\,56126 (IRAS\,07134+1005), the time-variable
H$\alpha$ profile usually contains an emission \citep{atlas1, Lebre}.
In the optical spectra of V510\,Pup (IRAS\,08005$-$2356), a powerful 
emission component is present both in the profiles of hydrogen lines, including 
the Paschen series lines, and in the selected lines of metal ions \citep{IRAS08005}.

The spectrum of the central star of the IR source RAFGL\,5081 is quite peculiar.
This star, like  GSC\,04050$-$02366, has an effective temperature
T$_{\rm eff}\approx5400$~K \citep{RAFGL5081} close to the evolutionary transition
from AGB to  post-AGB, according to \citet{Soker}.
In the spectrum of RAFGL\,5081, the  H$\alpha$ line has a  multicomponent and
time-variable profile, containing a weak emission \citep{RAFGL5081}. However,
as follows from  Fig.~\ref{Halpha}, H$\alpha$ profile in the spectra of  GSC\,04050$-$02366
is almost symmetrical, its absorption core is located near the average velocity 
Vr=+24.8\,km/s for all times of observation.

In the spectrum of GSC\,04050$-$02366 only forbidden emissions [O\,I]
$\lambda\,5577$, $\lambda\,6300$\,\AA\ of telluric origin are registered.
A weak emission raises the shortwave H$\alpha$ profile wing only for those two 
observation dates for which the  C$_2$ molecule Swan band is significantly 
weakened in Fig.~\ref{Swan}. This coincidence suggests the presence of asymmetry 
(or inhomogeneities) in the envelope.

The difference of radial velocities given in columns (2) and (4) of Table~\ref{velocity},
allows us to estimate the envelope expansion velocity, Vexp$\approx$16\,km/s, which agrees
with the result based on the CO profile \citep{Hrivnak99}.
If we consider the summary of Vexp values for post-AGB stars from \citet{Envelop},
we can see that the envelope expansion velocity in the IRAS\,Z02229 system is typical
for the objects of this type. Note here that most of the post-AGB stars with the
envelopes enriched in carbon have a similar envelope morphology
(bipolar or elongated halo) and an emission at $\lambda=21$~$\mu$m.

Comparison of  results of positional measurements in the spectra we obtained
on arbitrary dates of 2019$-$2021 revealed a variability of the heliocentric
radial velocity: averaged over the measurements of the positions of the set of
absorptions, the velocity varies in the range of Vr=$20.8\div26.1$\,km/s.
The presence of Vr variability is also confirmed by the earlier velocity value
Vr=18$\pm1$\,km/s, given in the paper of \citet{Reddy99} based on  December 1996 
observations.   The most probable cause of the radial velocity variability with 
the amplitude of about $\Delta$Vr$\approx 1.4$\,km/s can be the low-amplitude 
pulsations that are typical for post-AGB stars \citep{Aikawa}.
The star GSC\,04050$-$02366 with its fundamental parameters falls into the
region of Figure~1 in \citet{Aikawa}, populated with variable post-AGB stars.  
The presence of  pulsations in the atmosphere of this star is also confirmed 
by the photometric variability with the amplitude of $0\fm47$ based on the 
observations in the ASAS-SN database.

As a result of studying over the past $2$--$3$~decades of the candidates in post-AGB stars,
a group of objects with atmospheres enriched in  $s$-process heavy metals has been identified.
A complete list of stars of this type with a large set of data
(T$_{\rm_eff}$, luminosity, color excesses, SED, chemical composition of the atmosphere) is published
in the survey by \citet{Kamath}.  Counterparts of GSC\,04050$-$02366 based on the totality
of the observed properties (excess of carbon and heavy metals in the atmospheres,
a large color excess  owing to the powerful dust envelopes, spectral features of
carbon-containing molecules, emission at $\lambda=21$~$\mu$m) are the central
stars of the IR-sources  IRAS\,04296+3429, IRAS\,19500$-$1709, IRAS\,20000+3239,
IRAS\,22223+4327 and  IRAS\,23304+6147.  Post-AGB stars---the members of this 
group---when the main parameters are close, have their own specific differences as well.
For example, in the optical spectrum of the post-AGB star V448\,Lac in the IR source 
system IRAS\,22223+4327, all group indications are registered: radial velocity 
variability, manifestations of the circumstellar envelope in in the form of    
Swan system bands of the C$_2$ molecule, the asymmetry of strong absorptions of SiII, YII,
LaII, BaII ions \citep{V448Lac}.
At the same time, at certain dates  of observations  the spectrum of
V448\,Lac revealed an emission in the Swan band \citep{V448Lac}. V448\,Lac is also 
included in the reserved group of post-AGB stars with an emission at the wavelength 
of $21$~$\mu$m \citep{Volk1999}. This emission does not yet have a specific
identification, despite the many proposed versions of \citep{Zhang}.

The IR source IRAS\,04296+3429  system is also intriguing, which, similar to 
IRAS\,Z02229 has a powerful infrared flux.
The central star GSC\,02381$-$0104 in this  system has further
advanced to the planetary nebula stage than GSC\,04050$-$02366,
but in general, its fundamental parameters and energy distribution
in the spectrum of this system \citep{IRAS04296} are close to those
of the IR-source  IRAS\,Z02229 system.
The distant star GSC\,02381$-$0104 is faint for the high-resolution spectroscopy
(apparent brightness V$>14^{\rm m}$), but its optical spectrum (G8\,Ia) is close 
to that of  GSC\,04050$-$02366, while it holds the record for emission intensity
in the ($0;0$) and ($0;1$) Swan bands \citep{IRAS04296}. A greater farness
of this object from the AGB-phase probably leads to a significant expansion
of the enevelope, which contributes to the formation of emission
in the circumstellar details.

A nearby relative of GSC\,04050$-$02366 is a post-AGB supergiant V5112\,Sgr (IRAS\,19500$-$1709),
whose envelope is also enriched in heavy metals, which manifests itself in the 
splitting of resonance absorption profiles \citep{V5112Sgr}.
Along with this, the peculiarity of the optical spectrum of V5112\,Sgr is more pronounced: 
the profiles of strong BaII absorptions are split into three components. Two shortwave 
components origin in two different layers of the structured envelope of the star, 
formed at two previous stages of the evolution of the star and expanding with the 
velocities of Vexp$\approx20$ and 30\,km/s.

The cool post-AGB supergiant GSC\,04050$-$02366 belongs to the group of single
post-AGB stars that have passed the stage with the effective  $s$-process and the 
third mixing. This group currently includes about two dozen stars,
all of them are listed in the surveys by \citet{rev1, rev2, Kamath}. The splitting
of resonant absorptions of metals due to their removal to the circumstellar environment 
was found  in the spectra of only four of them.

\section{Main conclusions}\label{conclus}

The main new results of this study are as follows:
\begin{list}{}{
\setlength\leftmargin{2mm} \setlength\topsep{1mm}
\setlength\parsep{-0.5mm} \setlength\itemsep{2mm} }
\item{1) detection of a temporal  radial velocity variability based on the positional 
measurements of a great many of metal absorptions. The velocity Vr varies around the mean
value of Vr=24.75~km/s with the standard deviation from the mean of $\Delta$Vr=1.4\,km/s,
which may be a consequence of low-amplitude pulsations in the atmosphere;}
\item{2) detection of a temporal  variability  of the profiles of spectral features, 
including  the intensity of most of absorptions and Swan bands of the C$_2$ molecule, 
as well as a weak H$\alpha$ profile variability;}
\item{3) detection of splitting into two components (or asymmetries) of strong
low excitation absorptions (YII, ZrII,  BaII,  LaII,  NdII). The longwave component,
whose position is consistent with the position of symmetrical absorptions of other metals
is formed in the atmosphere of the star;}
\item{4) the position of the short-wave component of the split absorptions coincides with
the position of the Swan bands features, which points to its formation in the circumstellar
envelope expanding at a velocity of about Vexp=16\,km/s.
Therefore, the post-AGB supergiant GSC\,04050$-$02366 is the fourth object in
the group of stars for which the takeaway of the $s$-process heavy metals into the
circumstellar envelope was discovered}.
\end{list}

\section*{Acknowledges}
Observational data were obtained at the unique scientific facility, the 6-m Big 
Telescope Alt-azimuth of the SAO RAS. The study made use of the SIMBAD, VALD, SAO/NASA ADS,
ASAS-SN and Gaia~DR3 astronomical databases.

\section*{Funding}
The work on the spectroscopic data reduction and analysis of results
has been carried out within the framework of a grant of the
Ministry of Science and Higher Education of the Russian Federation
number~075-15-2022-262 (13.MNPMU.21.0003).

\section*{Conflict of interest}
The authors declare no conflict of interest.


\newpage

\begin{thebibliography}{39}
\providecommand{\natexlab}[1]{#1}

\bibitem[{Aikawa}(2010)]{Aikawa} T.~{Aikawa}, \aap\ \textbf{514}, A45 (2010).

\bibitem[{Akras} et~al.(2017)]{Akras} S.~{Akras}, J.~C.~{Ram{\'\i}rez V{\'e}lez}, N.~{Nanouris}, et~al., \mnras\
  \textbf{466}~(3), 2948 (2017).

\bibitem[{Barnbaum}(1994)]{atlas_Barn}  C.~{Barnbaum}, \apjs\ \textbf{90}, 317 (1994).

\bibitem[{Bloecker}(1995)]{Block1995}  T.~{Bloecker}, \aap\ \textbf{297}, 727 (1995).

 \bibitem[Brown et~al.(2021)]{Gaia} A.~G.~A.~{Brown}  et~al. (Gaia Collab.), \aap\
\textbf{649}, id.~A1 (2021).

\bibitem[{Di Criscienzo} et~al.(2016)]{Criscienzo} M.~{Di Criscienzo}, P.~{Ventura},
D.~A. {Garc{\'\i}a-Hern{\'a}ndez}, et~al.,  \mnras\ \textbf{462}~(1), 395 (2016).

\bibitem[{Groenewegen}(2022)]{Groen} M.~A.~T.~{Groenewegen}, \aap\ \textbf{659}, id.~A145 (2022).

\bibitem[{Herwig}(2005)]{Herwig2005} F.~{Herwig}, \araa\ \textbf{43}~(1), 435 (2005).

\bibitem[{H{\"o}fner} and {Olofsson}(2018)]{Hofner}
S.~{H{\"o}fner} and H.~{Olofsson}, Astron. Astroph. Rev. \textbf{26}~(1), id.~1 (2018).

\bibitem[{Hrivnak} and {Kwok}(1999)]{Hrivnak99} B.~J.~{Hrivnak} and S.~{Kwok}, \apj\ \textbf{513}~(2), 869 (1999).

\bibitem[{Hrivnak} et~al.(2010)]{Hrivnak2010}
B.~J. {Hrivnak}, W.~{Lu}, R.~E. {Maupin}, and B.~D.~{Spitzbart}, \apj\  \textbf{709}~(2), 1042 (2010).

\bibitem[{Kamath} et~al.(2022)]{Kamath} D.~{Kamath}, H.~{Van Winckel}, P.~{Ventura}, et~al., \apjl\ \textbf{927}~(1), id.~L13 (2022).

\bibitem[{Klochkova}(1997)]{rev1} V.~G.~{Klochkova}, \bsao\   \textbf{44}, 5 (1997).

\bibitem[{Klochkova}(2013)]{V5112Sgr} V.~G.~{Klochkova}, \alet\  \textbf{39}~(11), 765 (2013).

\bibitem[{Klochkova}(2014)]{Envelop} V.~G.~{Klochkova}, \ab\  \textbf{69}~(3), 279 (2014).

\bibitem[{Klochkova}(2019{\natexlab{a}})]{YHG} V.~G.~{Klochkova}, \ab\  \textbf{74}~(4), 475
  (2019{\natexlab{a}}).

\bibitem[{Klochkova}(2019{\natexlab{b}})]{rev2} V.~G.~{Klochkova}, \ab\  \textbf{74}~(4), 475
  (2019{\natexlab{b}}).

\bibitem[{Klochkova} and {Chentsov}(2004)]{IRAS08005} V.~G.~{Klochkova} and E.~L.~{Chentsov}, \arep\  \textbf{48}~(4), 301 (2004).

\bibitem[{Klochkova} et~al.(2017)]{RAFGL5081} V.~G.~Klochkova,  E.~L.~Chentsov, V.~E.~Panchuk,    N.~S.~Tavolzhanskaya, M.~V.~Yushkin,  \arep\   \textbf{61}, (11) 962  (2017).

\bibitem[{Klochkova} et~al.(2008)]{V2324Cyg} V.~G.~{Klochkova}, E.~L.~{Chentsov}, and V.~E.~{Panchuk}, \ab\  \textbf{63}~(2), 112 (2008).

\bibitem[{Klochkova} et~al.(2007{\natexlab{a}})]{atlas2} V.~G.~{Klochkova}, E.~L.~{Chentsov}, V.~E.~{Panchuk}, et~al., Baltic Astronomy \textbf{16}, 155 (2007{\natexlab{a}}).

\bibitem[{Klochkova} et~al.(2007{\natexlab{b}})]{atlas1} V.~G.~{Klochkova}, E.~L.~{Chentsov}, N.~S.~{Tavolganskaya}, and M.~V. {Shapovalov}, \ab\  \textbf{62}~(2), 162  (2007{\natexlab{b}}).

\bibitem[{Klochkova} et~al.(2010)]{V448Lac}
V.~G.~{Klochkova}, V.~E.~{Panchuk}, and N.~S.~{Tavolzhanskaya}, \arep\  \textbf{54}~(3), 234 (2010).

\bibitem[{Klochkova} et~al.(2022)]{SciNES} V.~G.~{Klochkova}, V.~E.~{Panchuk}, and M.~V.~{Yushkin}, \ab\
  \textbf{77}~(1), 84 (2022).

\bibitem[{Klochkova} et~al.(1999)]{IRAS04296} V.~G.~{Klochkova}, R.~{Szczerba}, V.~E.~{Panchuk}, and K.~{Volk}, \aap\ \textbf{345}, 905 (1999).

\bibitem[{Kochanek} et~al.(2017)]{Kochanek} C.~S.~{Kochanek}, B.~J.~{Shappee}, K.~Z.~{Stanek}, et~al., \pasp\
  \textbf{129}~(980), 104502 (2017).

\bibitem[{Kwok}(1993)]{Kwok93} S.~{Kwok}, \araa\ \textbf{31}, 63 (1993).

\bibitem[{Lebre} et~al.(1996)]{Lebre}  A.~{Lebre}, N.~{Mauron}, D.~{Gillet}, and D.~{Barthes}, \aap\ \textbf{310}, 923 (1996).

\bibitem[{Liu} et~al.(2018)]{Liu} N.~{Liu}, R.~{Gallino}, S.~{Cristallo}, et~al., \apj\ \textbf{865}~(2), 112
  (2018).

\bibitem[{Miroshnichenko} et~al.(2021)]{VES723} A.~S.~{Miroshnichenko}, V.~G.~{Klochkova}, E.~L.~{Chentsov}, et~al., \mnras \textbf{507}~(1), 879 (2021).

\bibitem[{Panchuk} et~al.(2017)]{NES} V.~E.~{Panchuk}, V.~G.~{Klochkova}, and M.~V.~{Yushkin}, \arep\
  \textbf{61}~(9), 820 (2017).

\bibitem[{Reddy} et~al.(1999)]{Reddy99} B.~E.~{Reddy}, E.~J.~{Bakker}, and B.~J.~{Hrivnak}, \apj\ \textbf{524}~(2), 831 (1999).

\bibitem[{Reyniers} et~al.(2002)]{Reyniers} M.~{Reyniers}, H.~{Van Winckel}, E.~{Bi{\'e}mont}, and P.~{Quinet}, \aap\ \textbf{395}, L35 (2002).

\bibitem[{Sahai} et~al.(2007)]{Sahai} R.~{Sahai}, M.~{Morris}, C.~{S{\'a}nchez Contreras}, and M.~{Claussen},
\aj\  \textbf{134}~(6), 2200 (2007).

\bibitem[{Soker}(2008)]{Soker} N.~{Soker},  \apjl\ \textbf{674}~(1), L49 (2008).

\bibitem[{Ueta} et~al.(2005)]{Ueta1} T.~{Ueta}, K.~{Murakawa}, and M.~{Meixner}, \aj\ \textbf{129}~(3), 1625 (2005).

\bibitem[{Ueta} et~al.(2007)]{Ueta2} T.~{Ueta}, K.~{Murakawa}, and M.~{Meixner}, \aj\ \textbf{133}~(4), 1345 (2007).

\bibitem[{Volk} et~al.(1999)]{Volk1999} K.~{Volk}, S.~{Kwok}, and B.~J.~{Hrivnak},  \apjl\ \textbf{516}~(2), L99 (1999).

\bibitem[{Zhang} et~al.(2009)]{Zhang} K.~{Zhang}, B.~W.~{Jiang}, and A.~{Li}, \mnras\ \textbf{396}~(3), 1247 (2009).

\end{thebibliography}
\end{document}